\documentclass[twocolumn,twoside]{IEEEtran}
\usepackage{times}
\usepackage{amsmath}  
\usepackage{amssymb}  
\usepackage{mathrsfs} 
\usepackage{theorem}  
\usepackage{cite}     
\usepackage{comment}  
\usepackage{upref}
\usepackage{amsfonts}
\usepackage{verbatim}
\usepackage[dvipsnames,usenames]{color}
\usepackage{enumerate}
\usepackage{graphicx}
\usepackage{subfigure}
\usepackage{latexsym}
\usepackage{caption}
\usepackage{algorithm}
\usepackage[noend]{algorithmic}
\usepackage{psfrag}
\usepackage[normalem]{ulem}

\usepackage{color}
\usepackage{times,color}
\usepackage[usenames,dvipsnames,svgnames,table]{xcolor}

\usepackage[top=0.58in, bottom=0.58in, left=0.6in, right=0.6in]{geometry}

\newcommand{\code}{\ttfamily\bfseries}

\newcommand{\be}[1]{\begin{equation}\label{#1}}
\newcommand{\ee}{\end{equation}}

\newcommand{\bc}{\begin{center}}
\newcommand{\ec}{\end{center}}

\newcommand{\ceil}[1]{\lceil{#1}\rceil}

\newcommand{\qed}{\hfill$\Box$\\[1ex]}

\newcommand{\bfc}{{\boldsymbol c}}

\newcommand{\bfp}{{\boldsymbol p}}

\newcommand{\bfu}{{\boldsymbol u}}
\newcommand{\bfv}{{\boldsymbol v}}
\newcommand{\bfw}{{\boldsymbol w}}

\renewcommand{\le}{\leqslant}
\renewcommand{\leq}{\leqslant}
\renewcommand{\ge}{\geqslant}
\renewcommand{\geq}{\geqslant}

\newcommand{\F}{\mathbb{F}}

\newcommand{\R}{\mathbb{R}}

\newcommand{\C}{\mathbb{C}}

\newcommand{\Cref}[1]{Co\-rol\-la\-ry\,\ref{#1}}

\theoremstyle{plain} \theorembodyfont{\normalfont\slshape}

\newtheorem{thm}{Theorem$\!$}
\newenvironment{theorem}{\begin{thm}\hspace*{-1ex}{\bf.}}{\end{thm}}

\newtheorem{prop}[thm]{Proposition$\!$}

\newtheorem{lem}[thm]{Lemma$\!$}
\newenvironment{lemma}{\begin{lem}\hspace*{-1ex}{\bf.}}{\end{lem}}

\newtheorem{cor}[thm]{Corollary$\!$}
\newenvironment{corollary}{\begin{cor}\hspace*{-1ex}{\bf.}}{\end{cor}}

\newtheorem{const}[thm]{Construction$\!$}
\newenvironment{construction}{\begin{const}\hspace*{-1ex}{\bf.}}{\end{const}}

\newtheorem{proc}[thm]{Procedure$\!$}

\newtheorem{cl}[thm]{Claim$\!$}

\newtheorem{conject}[thm]{Conjecture$\!$}

\newtheorem{defi}[thm]{Definition$\!$}
\newenvironment{definition}{\begin{defi}\hspace*{-1ex}{\bf.}}{\end{defi}}

\theorembodyfont{\normalfont}

\newtheorem{exam}{Example$\!$}
\newenvironment{example}{\begin{exam}\hspace*{-1ex}{\bf .}}{\qed\end{exam}}

\newtheorem{remrk}{Remark$\!$}

\definecolor{Codecolor}{named}{White}  

\newcommand{\Copen}{\mbox{\{\kern-5.50pt\{}}
\newcommand{\Cclose}{\mbox{\}\kern-5.50pt\}}}
\newcommand{\Cslash}{\mbox{$\backslash\kern-6.02pt\backslash$}}

\newcommand{\FF}{\mathbb F}

\begin{document}
\title{\textbf{\huge{
\hspace*{-1pt}Coding for Racetrack Memories}\hspace*{-1pt}}
}
\author{\textbf{Yeow Meng Chee},\!\IEEEauthorrefmark{1}
        \textbf{Han Mao Kiah},\!\IEEEauthorrefmark{1}
        \textbf{Alexander Vardy},\!\IEEEauthorrefmark{2}
        \textbf{Van Khu Vu},\!\IEEEauthorrefmark{1}
        and \textbf{Eitan Yaakobi}\IEEEauthorrefmark{3}

\IEEEauthorblockA{\IEEEauthorrefmark{1} 
Nanyang Technological University, Singapore\\}
\IEEEauthorblockA{\IEEEauthorrefmark{2} 
University of California San Diego, La Jolla, CA\,92093, USA \\}
\IEEEauthorblockA{\IEEEauthorrefmark{3} 
Technion --- Israel Institute of Technology, Haifa, 32000 Israel\\}
{Emails:\code  \{ymchee,hmkiah,vankhu001\}@ntu.edu.sg.edu},\,
\code  avardy@ucsd.edu,\,
\code  yaakobi@cs.technion.ac.il\vspace{-5ex}} \vspace{-3ex}
\maketitle

\thispagestyle{empty}
\pagestyle{empty}
\begin{abstract}
\emph{Racetrack memory} is a new technology which utilizes magnetic domains along a nanoscopic wire in order to obtain extremely high storage density. In racetrack memory, each magnetic domain can store a single bit of information, which can be sensed by a reading \emph{port} (\emph{head}). The memory has a tape-like structure which supports a \emph{shift} operation that moves the domains to be read sequentially by the head. In order to increase the memory's speed, prior work studied how to minimize the latency of the shift operation, while the no less important reliability of this operation has received only a little attention. 

In this work we design codes which combat shift errors in racetrack memory, called \emph{position errors}. Namely, shifting the domains is not an error-free operation and the domains may be over-shifted or are not shifted, which can be modeled as \emph{deletions} and \emph{sticky insertions}. While it is possible to use conventional deletion and insertion-correcting codes, we tackle this problem with the special structure of racetrack memory, where the domains can be read by multiple heads. Each head outputs a noisy version of the stored data and the multiple outputs are combined in order to reconstruct the data. Under this paradigm, we will show that it is possible to correct, with at most a single bit of redundancy, $d$ deletions with $d+1$ heads if the heads are well-separated. Similar results are provided for burst of deletions, sticky insertions and combinations of both deletions and sticky insertions.
\end{abstract}

\vspace{-3ex}
\section{Introduction}
\emph{Racetrack memory}, also known as \emph{domain wall memory}, is an emerging non-volatile memory which is based on spintronics technology. It attracts significant attention due to its promising ultra-high storage density, even comparing to other spintronics memory technologies such as STT-RAM~\cite{Zhangetal15}. 

A racetrack memory is composed of \emph{cells}, also called \emph{domains}, which are positioned on a tape-like stripe and are separated by \emph{domain walls}. The magnetization of a domain is programmed to store a single bit value, which can be read by sensing its magnetization direction. The reading mechanism is operated by a read-only \emph{port}, called a \emph{head}, together with a \emph{reference domain}. Since the head is fixed (i.e. cannot move), a \emph{shift} operation is required in order to read all the domains. Shifting the cells is accomplished by applying a shift current which moves the domain walls in one direction. Thus, shift operations move all the domains one step either to the right or to the left. It is also possible to shift by more than a single step by applying a stronger current. When doing so, it is required to have more than a single head to read the domain walls~\cite{PHT08}.

There are several approaches to enhance the shift operation in order to reduce its time and energy consumption\cite{SWL13, V_etal12}. However these mechanisms suffer from degraded reliability and cannot ensure that domains are perfectly shifted so they are aligned with the head. These errors, called \emph{position errors}, can be modeled as deletions and sticky insertions~\cite{Zhangetal15}, which is the motivation for this work. A deletion is the event where the domains are shifted by more than a single domain location and thus one of the domains is not read, which results with a \emph{deletion} of the bit stored in this domain. In case the domains were not successfully shifted, then the same domain is read again and we experience an \emph{insertion}, however of the same bit. This kind of insertion errors is also referred as \emph{repetition errors} or \emph{sticky insertions} in a \emph{sticky channel}~\cite{DA10,MV17}. 

In this work we study codes which correct position errors in racetrack memory. At first sight, this problem is not any different than the well-studied problem of designing codes correcting deletions and insertions~\cite{BGZ15,Lev66}. However, we take another approach to tackle the problem and leverage the special features of racetrack memory, where it is possible to use more than a single head in order to read the domains. Thus, each domain is read more than once and the extra reads can be used in order to correct the position errors during the reading process. Since every head reads all the bits, we can treat every head as a channel which returns a noisy version of the stored information, and based on these noisy reads the information is decoded. This model falls under the general framework by Levenshtein of the \emph{reconstruction problem}~\cite{L01A}. However, in our case, as opposed to the general one studied by Levenshtein, the position errors are correlated and depend on the locations and distance between the different heads. 

In contrast to substitution errors, deletions/sticky insertions behave {\em differentially}.
Namely, to successfully decode a substitution error, it is necessary to determine the location of the error. 
However, for deletions/sticky insertions, the decoder can successfully decode the correct codeword without determining all the locations of the deletions/sticky insertions, since it could be any bit which belongs to the run where each deletion/sticky insertion has occurred. Assume first that the heads are adjacent and on every cycle the domains are shifted by a single location. Thus, if there are no position errors, the bit stored in each domain is read twice. On the other hand, in the occurrence of position errors, the deletions/sticky insertions in the two heads are correlated. For example, if the $i$th bit is deleted in the first head then the $(i+1)$-st bit is deleted in the second head. In case these two deleted bits belong to the same run, then the noisy words from the two heads are identical and thus we did not benefit from the extra read by the additional head. On the hand, if the heads are well separated and there are no long runs in the stored information, then the heads' outputs will differ and under this setup we will show how it is possible to correct the position errors. Note that it is possible to correct a fixed number of deletions and sticky insertions with a single head while the rate of the codes approaches 1 and the redundancy order is $\Theta(\log(n))$~\cite{BGZ15,Lev66}. Hence, any code construction using multiple heads should have rate approaching 1 and more than that, improve the redundancy result of $\Theta(\log(n))$. However, this should be accomplished while minimizing the distance between the heads. 

The rest of this paper is organized as follows. In Section~\ref{sec:defs}, we formally define the model and problems studied in the paper, namely the reading process in racetrack memory and codes correcting deletions and sticky insertions using multiple heads. In Section~\ref{sec:2H1D}, we construct codes correcting a single deletion using two heads with approximately 0.36 redundancy bits, by requiring the distance between the heads to be at least $\ceil{\log(n)}+1$. In Section~\ref{sec:burst}, we extend this construction for codes correcting a burst of deletions where the length of the burst is either exactly $b$ or at most $b$. Another extension is given in Section~\ref{sec:multiple deletions} for codes correcting multiple deletions. In this case our construction can correct $d$ deletions using $d+1$ heads with at most a single bit of redundancy, by requiring the distance between adjacent heads to be at least $d\ceil{\log(n)}+d(d+1)/2+1$. In the case the number of heads $m$ is strictly less than $d+1$, we show that it is possible to correct $m-1$ deletions with the heads, and so the code should be able to correct the remaining $d-(m-1)$ deletions. In this section, we also report on several more results we could not include due to the lack space. Some of the proofs are omitted for the same reason. 
\vspace{-1.5ex}
\section{Preliminaries and Model Definitions}\label{sec:defs}
Let $\FF_2$ denote the binary field. For a positive integer $n$, the set $\{1,2,\ldots,n\}$ is denoted by $[n]$. Let $\bfu=(u_1,\ldots,u_n)$ and $\bfv=(v_1,\ldots,v_m)$ be two vectors of length $n$ and $m$, respectively. The concatenation of $\bfu$ and $\bfv$ is the vector $(u_1,\ldots,u_n,v_1,\ldots,v_m)$ of length $n+m$, which is denoted by $\bfu \circ \bfv$. A subvector of a word $\bfu$ is a vector $\bfu[i_1,i_2]=(u_{i_1},u_{i_1+1},\ldots,u_{i_2}) \in \FF_2^n$ in which $1 \leq i_1 \leq i_2 \leq n$. The length of this subvector is $1 \leq i_2-i_1+1 \leq n$. In case $i_1=i_2=i,$ we denote a subvector $\bfu[i,i]$ of length 1 by $\bfu[i]$ to specify the $i$-th element of vector $\bfu$.

Let $\ell$ and $m$ be two positive integers where $\ell\leq m$. Then, a length-$m$ vector $\bfv\in \FF_2^m$ which satisfies $v_i=v_{i+\ell}$ for all $1\leq i\leq m-\ell$ is said to have \emph{period $\ell$}. For a vector $\bfu \in \FF_2^n$, we denote by $L(\bfu,\ell)$ the length of its longest subvector which has period $\ell$. Note that by definition $L(\bfu,\ell)\geq \ell$, and for $\ell=1$, $L(\bfu,1)$ equals the length of the longest run in $\bfu$. \vspace{-2ex}
\begin{example}
Let $\bfu=(u_1,\dots,u_9)=(0,0,1,1,0,1,0,1,1) \in \FF_2^9$. Since the longest run in $\bfu$ is of length two, we have $L(\bfu,1)=2$. The subvector $\bfu[4,8]=(1,0,1,0,1)$ of $\bfu$ has period 2 since $u_4=u_6=u_8=1$ and $u_5=u_7=0.$ This is the longest subvector of $\bfu$ of period $2$, and hence $L(\bfu,2)=5$.
\end{example}\vspace{-4ex}

For a length-$n$ word $\bfu\in \FF_2^n$ and $i\in[n]$, we denote by $\bfu(\delta_i)$ the vector obtained by $\bfu$ after deleting its $i$th bit, that is, $\bfu(\delta_i) = (u_1,\ldots, u_{i-1},u_{i+1},\ldots,u_n)$. For a set $\Delta \subseteq \{\delta_i: i \in [n]\}$, we denote by $\bfu(\Delta)$ the vector of length $n-|\Delta|$ obtained from $\bfu$ after deleting all the bits specified by the locations in the set $\Delta$. In case $\Delta= \{\delta_i,\ldots,\delta_{i+b-1}\}$ then we denote the vector $\bfu(\Delta)$ by $\bfu(\delta_{[i,b]})$ to specify a burst of $b$ deletions starting at the $i$th position.
\vspace{-2ex}
\begin{example}
Let $\bfu=(0,0,1,1,0,1,0,1,1) \in \FF_2^9$, then $\bfu(\delta_4)=(0,0,1,0,1,0,1,1).$ For $\Delta=\{\delta_4, \delta_7,\delta_9\}$ then $\bfu(\Delta)=(0,0,1,0,1,1)$, and $\bfu(\delta_{[3,4]}) = (0,0,0,1,1)$.
\end{example}\vspace{-4ex}

\begin{figure}[h]
 \centering
 \captionsetup{justification=centering}
  \includegraphics[width=0.5\textwidth]{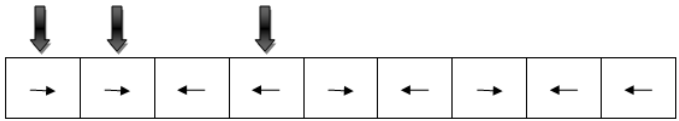}
   \caption {Racetrack memory with multiple heads}
    \label{fig:racetrack1}\vspace{-2ex}
\end{figure}

In this work, we assume that the information stored in the racetrack memory is represented by a word $\bfu$. The memory is comprised of magnetizable cells which can store a single bit. The information is read back from the cells by sensing their magnetization direction using heads which are fixed in their positions; see Fig.~\ref{fig:racetrack1}. \vspace{-2ex}
Since the heads are fixed in their locations, the memory cells move so they can all be read by the heads. This \emph{shifting operation} is performed by applying a shift current which moves all the cells on each cycle one or more steps in the same direction~\cite{PHT08}. However, the shifting mechanism does not work perfectly and may suffer from errors, called \emph{position errors}. That is, cells may be shifted by more than a single location on each cycle or are not shifted. These position errors can be modeled as deletions and sticky insertions. Namely, a \emph{single deletion} is the event where the cells are shifted by two locations instead of one and thus one of the bits is not read by the head. In case the cells were shifted by some $b+1>2$ locations, then $b$ consecutive cells were not read and we say that a \emph{deletion burst of size $b$} has occurred. On the other hand, a \emph{sticky insertion} is the event where the cells were \emph{not} shifted and the same cell is read again and if this happens $b>1$ times in a row, we say that a \emph{burst of $b$ sticky insertions} has occurred. 

We assume that there are several heads and each head reads \emph{all} the cells. In case there is only a single head, then the only approach to correct the position errors is by using a code which is capable of correcting deletions and sticky insertions. However, in case there are several heads, the cells are read multiple times by each head and thus we study how this inherent redundancy can be used to design better codes. The output of the heads depend on their locations. For example, assume there are three heads which are used to read the stored word $\bfu$, where the distance between the first two heads is $t_1$ and the distance between the last two heads is $t_2$. If a deletion occurs at position $i$ in the first head then a deletion also occurs at position $i+t_1$ in the second head and another deletion at position $i+t_1+t_2$ in the third head. Therefore, the output of the first, second, third head is the vector $\bfu(\delta_i),\bfu(\delta_{i+t_1}), \bfu(\delta_{i+t_1+t_2})$, respectively.

The goal in this paper is to design codes correcting position errors in the reading process. We say that a code is an \emph{$m$-head $b$-position-error-correcting code} if it can correct $b$ position errors using $m$ heads. Similarly, we also define \emph{$m$-head $b$-deletion-correcting codes,} \emph{$m$-head $b$-sticky-insertion-correcting codes}, \emph{$m$-head $b$-burst-deletion-correcting codes}, and \emph{$m$-head $b$-burst-sticky-insertion-correcting codes}. We note that the locations of the heads is also part of the code design. Since the area for shifting the cells is constrained, the heads should not be too far apart and the distance between adjacent heads should thus be minimized. As always, the goal in designing these codes is to minimize the redundancy of each code construction.

\section{Two-Head Single-Deletion-Correcting Codes}\label{sec:2H1D}
In this section we study how to construct two-head single-deletion-correcting codes. Our main result states that if the distance between the two heads is at least $\ceil{\log(n)}+1$ cells, where $n$ is the length of the codewords, then such codes exist with redundancy of roughly $0.36$ bits.\vspace{-2ex} 
\begin{construction}
For all $t\leq n$, let $\C_1(n,1,t)$ be a code of length $n$ such that the length of the longest run of every codeword is at most $t$. That is, $\C_1(n,1,t)=\{\bfc \in \FF_2^n \ | \ L(\bfc,1) \leq t\}$.
\end{construction}

The following theorem proves the correctness of this construction.
\begin{theorem}\label{thm1.1}
The code $\C_1(n,1,t)$ is a two-head single-deletion-correcting code when the heads are positioned $t$ locations apart. 
\end{theorem}
\begin{IEEEproof}
Let $\bfc=(c_1,\dots,c_n) \in \C_1(n,1,t)$ be a stored codeword of length $n$ and assume that a single deletion occurred at position $i$. Then, the outputs from the two heads are:
\begin{align*}
\textbf{Head 1: } \bfc(\delta_i)&=(c_1,\dots,c_{i-1},c_{i+1},\dots,c_n),\\
\textbf{Head 2: } \bfc(\delta_{i+t})&=(c_1,\dots,c_{i+t-1},c_{i+t+1},\dots,c_n).
\end{align*}
Consider the first $i+t-1$ bits in these two sequences:
\begin{align*}
\hspace{-1ex}\textbf{Head 1: }\hspace{-0.4ex} \bfc(\delta_i)[1,i+t-1]&=(c_1,\dots,c_{i-1},c_{i+1},c_{i+2},\dots,c_{i+t}),\\
\hspace{-1ex}\textbf{Head 2: }\hspace{-0.4ex} \bfc(\delta_{i\hspace{0ex}+\hspace{0ex}t})[1,i\hspace{-0.4ex}+\hspace{-0.4ex}t\hspace{-0.4ex}-\hspace{-0.4ex}1]&\hspace{-0.4ex}=\hspace{-0.4ex}(c_1,\dots,c_{i-1},c_{i},c_{i+1},\dots,c_{i+t-1}).
\end{align*}
We claim that $\bfc(\delta_i)[1,i+t-1] \neq \bfc(\delta_{i+t})[1,i+t-1]$. Otherwise, we will get that \vspace{-1ex}
$$c_i=c_{i+1}=\cdots = c_{i+t-1} = c_{i+t},\vspace{-1ex}$$ 
which implies that there is a run of length $t+1$ in $\bfc$ in contradiction to the construction of the code $\C_1(n,1,t)$. Let $j_1$ be the leftmost index that differs between $\bfc(\delta_i)[1,i+t-1]$ and $\bfc(\delta_{i+t})[1,i+t-1]$. Such an index exists since $\bfc(\delta_i)[1,i+t-1] \neq \bfc(\delta_{i+t})[1,i+t-1]$ and so $j_1 \leq i+t-1$. Furthermore, $j_1 \geq i$ since the first $i-1$ bits in the outputs from two heads are the same as in the stored codeword. Note that $j_1$ can be different than $i$ in case the $i$th bit which was deleted is in a middle of a run and so the first occurrence where $\bfc(\delta_i)[1,i+t-1]$ and $\bfc(\delta_{i+t})[1,i+t-1]$ differ is only at the end of this run. We conclude that $\bfc[1,j_1]=\bfc(\delta_{i+t})[1,j_1]$ and $\bfc[j_1+1,n]=\bfc(\delta_i)[j_1,n-1]$. Hence, the original codeword $\bfc$ can be recovered by concatenating the first $j_1$ bits from $\bfc(\delta_{i+t})$ and the last $n-j_1$ bits from $\bfc(\delta_i)$. That is, $\bfc = \bfc(\delta_{i+t})[1,j_1] \circ \bfc(\delta_i)[j_1,n-1]$. This proof also provides a simple decoding algorithm for the code $\C_1(n,1,t)$.
\end{IEEEproof}

The next example demonstrates this code construction and its decoder.\vspace{-2ex}
\begin{example}
Let $n=9,t=3$ and $\bfc=(0,0,1,1,0,1,0,1,1)$ be a stored codeword in $\C_1(n,1,t).$
Let us assume that the outputs from two heads are:
\begin{align*}
\textbf{Head 1: } \bfc(\delta_3)&=(0,0,1,0,1,0,1,1),\\
\textbf{Head 2: } \bfc(\delta_{6})&=(0,0,1,1,0,0,1,1).
\end{align*}
Hence, $j_1=4$ is the leftmost index that differs between two vectors and thus the stored codeword is decoded according to $\bfc=\bfc(\delta_6)[1,4]\circ \bfc(\delta_3)[4,8]= (0,0,1,1,0,1,0,1,1).$
\end{example}\vspace{-2ex}

By a suitable mapping described in Section IV, the code $\C_1(n,1,t)$ can be transformed into a code that satisfies the $(0,t-1)$ \emph{Run Length Limited} (\emph{RLL}) constraint~\cite{Immink04}. While efficient encoding and decoding algorithms are known for codes which satisfy the $(0,t-1)$ RLL constraint for fixed value of $t$, the rates of these codes is strictly less than 1. Since we can achieve codes with rate approaching 1 by simply using a single head and a single-deletion-correcting code of redundancy at most $\log(n+1)$~\cite{Lev66}, we are interested only in codes with rate approaching 1 and will  optimize their redundancy. Thus, we follow a similar approach to the one taken in~\cite{SWBY16} for codes correcting a burst of deletions and let $t$ be a function of the code length $n$. In particular, by choosing $t=\ceil{ \log(n)}+1$, it was observed in~\cite{SWBY16}, using the derivations from~\cite{S90} and~\cite{S12}, that the redundancy of the code $\C_1(n,1,\lceil\log(n)\rceil+1)$ is approximately 0.36, and for $t=\ceil{\log(n)}+2$ efficient encoding and decoding algorithms were recently found for these codes using a single bit of redundancy~\cite{LY17}. We conclude this discussion with the following corollary.\vspace{-2ex}
\begin{corollary}\label{cor:2head1deletion}
There exists a two-head single-deletion-correcting code when the heads are positioned $t=\ceil{\log(n)}+1$ locations apart with redundancy of approximately $\log(e)/4\approx 0.36$ bits.
\end{corollary}\vspace{-3ex}

\section{Codes Correcting a Burst of Deletions}\label{sec:burst}

In this section we study the setup where the domains are over-shifted by more than a single location, so a burst of deletions occurs in each head. We will focus on two cases: the length of the burst is exactly $b$ or at most $b$.

\vspace{-2ex}
\subsection{Two-Head $b$-Burst-Deletion-Correcting Codes}

Here we investigate codes correcting a burst of exactly $b$ adjacent deletions using two heads. Suppose we use two heads at distance $t$ to correct a burst of size $b$ in the stored codeword $\bfc=(c_1,\dots,c_n)$.
Recall that for $i \in [n]$ and $b \in [n-i]$, the vector obtained from $\bfc$ after deleting the subvector $c[i,i+b-1]=(c_i,\dots,c_{i+b-1})$ is $\bfc(\delta_{[i,b]})$.
Therefore, we know that if the output from the first head is $\bfc(\delta_{[i,b]})$ for some $i$ and $b$, then the output from the second head is $\bfc(\delta_{[i+t,b]})$, where the heads are located $t$ positions apart. The following is the construction of such codes.\vspace{-2ex}
\begin{construction}
Let $\C_2(n,b,t)$ be a code of length $n$ such that the length of the longest subvector which has period $b$ of every codeword $\bfc \in \C_2(n,b,t)$ is at most $t$. That is, $\C_2(n,b,t)=\{\bfc \in \F_2^n \ | \ L(\bfc,b) \leq t\}$.
\end{construction}
The proof that this construction can correct a burst of deletion of length $b$ follows similar ideas from the proof of Theorem~\ref{thm1.1}.
\begin{theorem}\label{th:burst}
The code $\C_2(n,b,t)$ is a two-head $b$-burst-deletion-correcting code when the heads are positioned $t$ locations apart.
\end{theorem}

Next we turn to evaluate the size of the code $\C_2(n,b,t)$. In particular, as done in the previous section, we will find a value of $t$ for which the redundancy of the code will be approximately $0.36$ bits. Let us start with the following definition.
\begin{definition} 
Let $\bfu=(u_1,\ldots, u_m)\in \F_2^m$ be a length-$m$ binary vector. For $b>m$, the \emph{$b$-period check vector} of $\bfu$ is 
the vector $\bfp_b(\bfu)=(u_1+u_{1+b},\ldots, u_{m-b}+u_{m})\in \F_2^{m-b}$ of length $m-b$.
\end{definition}
The following lemma can be readily verified.
\begin{lemma}\label{lem:runs}
A word $\bfu$ contains a subvector of length $t$ with period $b$  if and only if  $\bfp_b(\bfu)$ contains a run of $t-b$ zeroes.
\end{lemma}
For a vector $\bfu$, we denote by $L_0(\bfu)$ the length of the longest run of zeroes in $\bfu$. For example $L_0(0110100010) =3$. For all $n$ and $t\leq n$, we define the code $\R(n,t)$ to be 
$$\R(n,t) = \{  \bfc\in \F_2^n \ | L_0(\bfu) \leq t\}.$$
Using Lemma \ref{lem:runs}, we can construct a bijection between $\C_2(n,b,t)$
and the set $\F_2^b\times \R(n-b,t-b)$ for $n\ge b+1$. Specifically, we define the following maps.
\begin{itemize}
\item $\Phi:\C_2(n,b,t)\to \F_2^b\times \R(n-b,t-b)$, where $\Phi(\bfu)=(\bfu[1,b],\bfp_b(\bfu))$.
\item $\Psi: \F_2^b\times \R(n-b,t-b)\to \C_2(n,b,t)$, where $\Psi(\bfv,\bfw)=\bfu$ and
\[
u_i=
\begin{cases}
v_i, & \mbox{if } i\le b,\\
u_{i-b}+w_{i-b}, & \mbox{otherwise}.
\end{cases}
\]
\end{itemize}

In the context of error-correcting codes for tandem duplications \cite{JFSB},
Jain {\em et al.} demonstrated Lemma \ref{lem:runs} and the fact that $\Phi$ and $\Psi$ are bijections 
when $t=2b-1$. It is straightforward to extend the proof for $t\ge b$.
Hence, we have the following lemma that is useful in evaluating the size of the code $\C_2(n,b,t)$.
\begin{lemma}\label{lem:size}
For all $n,b,t$, $|\C_2(n,b,t) |  = 2^b\cdot |\R(n-b,t-b)|.$
\end{lemma}

The size of the code $\R(n,t)$ can be calculated using the results from Section~\ref{sec:2H1D} and by applying Lemma~\ref{lem:size} for $b=1$ to get that for all $n$ and $t\leq n$,
$ |\R(n,t)| = {|\C_1(n+1,1,t+1)|}/2.$ We can now conclude with the following corollary.\vspace{-2ex}
\begin{corollary} \label{cor:codesize1}
For all $n,b,t$,
$$|\C_2(n,b,t) |  = 2^b\cdot \frac{|\C_1(n-b+1,1,t-b+1)|}{2}.$$
\end{corollary}\vspace{-0ex}
According to Corollaries~\ref{cor:2head1deletion} and~\ref{cor:codesize1} we conclude the following.\vspace{-2ex}
\begin{corollary}\label{cor:2headburstdeletion}
There exists a two-head $b$-burst-deletion-correcting code when the heads are positioned $t=\ceil{\log(n)}+b$ locations apart with redundancy of approximately $\log(e)/4\approx 0.36$ bits.
\end{corollary}\vspace{-4ex}

\subsection{Correcting a Burst of Length at Most b}

The goal of this section is to design a code correcting a burst of at most $b$ deletions using two heads. We follow the same ideas presented thus far and use the following construction.\vspace{-1ex}
\begin{construction}
Let $\C_3(n,\leq b,t)$ be a code of length $n$ which is the intersection of the codes $\C_2(n,\ell,t)$ for $1\leq \ell \leq b.$
That is,
\begin{align*}
\C_3(n,\leq b,t)& = \cap_{\ell=1}^{b} \C_2(n,\ell,t) & \\
& =\{ \bfc \in \F_2^n \ | \ L(\bfc,\ell) \leq t, \text{ for all } \ell \leq b \}. &
\end{align*}
\end{construction}
\vspace{-2ex}
\begin{theorem}
The code $\C_3(n,\leq b, t)$ can correct up to $b$ consecutive deletions using two heads at distance $t$.
\end{theorem}

In this case we will not able to provide an exact approximation for the redundancy of the code $\C_3(n,\leq b,t)$ as in previous cases. However, we will find a value of $t$ for which the redundancy of the code is at most a single bit. For this purpose, we follow similar ideas to the ones presented by Schoeny \emph{et al.} in \cite{SWBY16} when studying the redundancy of the so-called \emph{universal RLL constraint}. This result is stated in the next theorem.\vspace{-1ex}
\begin{theorem}
For all $n,b,t$, \vspace{-1.5ex}
$$|\C_3(n,\leq b,t) | \geq  2^n\left(1- n \cdot \left( \frac{1}{2} \right)^{t-b}   \right).\vspace{-1ex}$$
In particular, for $t=\ceil{\log(n)}+b+1$ the redundancy of the code $\C_3(n,\leq b,t)$ is at most a single bit.
\end{theorem}\vspace{-3ex}

\section{Codes Correcting Multiple Deletions}\label{sec:multiple deletions}

In this section we move to the more challenging task of correcting multiple deletions and construct $m$-head $d$-deletion-correcting codes. For simplification, we first consider the case $d=2$ and show that the code $\C_3(n,\leq 2,t_1)$, which can correct a burst of at most two deletions by using two heads, is a three-head double-deletion-correcting code, when the distance between every adjacent heads is at least $t=2(t_1-1)$. We will then use this result as a building block for a more general claim on codes which can correct $d$ deletions using $m$ heads. While we don't design new code constructions, a key point in the construction is finding the required minimum distance between two adjacent heads for its success.

We start by presenting our result for the construction of three-head double-deletion-correcting codes. \vspace{-1ex}
\begin{theorem}\label{multideletion:thm1}
The code $\C_3(n,\leq 2,t_1)$ is a three-head double-deletion-correcting code when the distance between adjacent heads is at least $t=2(t_1-1)$.
\end{theorem}

\begin{IEEEproof}
Let $\bfc=(c_1,\dots,c_n) \in \C_3(n,\leq 2, t_1)$ be the stored codeword and $t=2(t_1-1)$ be the distance between adjacent heads.
Let us assume that the two deletions occurred in the first head are in positions $i_1,i_2$, where $i_1<i_2$. Hence the deletions in the second head are in positions $i_1+t, i_2+t$ and in the third head they are in positions $i_1+2t, i_2+2t$. That is, the outputs from the three heads are:
\begin{align*}
&\textbf{Head 1: }  \bfc(\delta_{ i_1},\delta_{ i_2}) \\
&=(c_1,\dots,c_{i_1-1},c_{i_1+1},\dots,c_{i_2-1},c_{i_2+1},\dots,c_n),\\
&\textbf{Head 2: } \bfc(\delta_{i_1+t} ,\delta_{ i_2+t}) & \\ 
& =(c_1,\ldots,c_{i_1+t-1},c_{i_1+t+1},\dots,c_{i_2+t-1},c_{i_2+t+1},\ldots,c_n),\\
&\textbf{Head 3: } \bfc(\delta_{i_1+2t},\delta_{i_2+2t})&  \\
&=(c_1,\ldots,c_{i_1+2t-1},c_{i_1+2t+1},\dots,c_{i_2+2t-1},c_{i_2+2t+1},\ldots,c_n).
\end{align*}

We prove that it is possible to correct the two deletions by explicitly showing how to decode them. This will be done in three steps. 
\begin{enumerate}
\item First, use the first two heads to correct the first deletion in the first head. 
\item Then, use the second and third heads to correct the first deletion in the second head. 
\item At this point, the first and second heads have only a single deletion and thus we proceed to correct this deletion as was done in Theorem~ \ref{thm1.1}.
\end{enumerate}
Since the last two steps are very similar to the first one, we only discuss the first step. 

In order to prove the first step, we show that $\bfc(\delta_{ i_1},\delta_{ i_2}) [1,i_1+t-1] \neq \bfc(\delta_{ i_1+t},\delta_{ i_2+t}) [1,i_1+t-1].$ Assume in the contrary, then we distinguish between the following two cases:
\begin{itemize}
\item Case 1: If $i_2-i_1 \geq t_1+1$ then the two subvectors
\begin{align*}
&\hspace{-0.6ex}\bfc(\delta_{ i_1},\delta_{ i_2})[1,i_1\hspace{-0.4ex}+\hspace{-0.4ex}t_1\hspace{-0.4ex}-\hspace{-0.4ex}1]\hspace{-0.4ex}=\hspace{-0.4ex}(c_1,\dots,c_{i_1-1},c_{i_1+1},\dots,c_{i_1+t_1}),\\
&\hspace{-0.6ex}\bfc(\delta_{ i_1\hspace{-0.1ex}+\hspace{-0.1ex}t},\delta_{ i_2\hspace{-0.1ex}+\hspace{-0.1ex}t})[1,i_1\hspace{-0.5ex}+\hspace{-0.5ex}t_1\hspace{-0.5ex}-\hspace{-0.5ex}1]\hspace{-0.5ex}=\hspace{-0.5ex}(c_1,\hspace{-0.2ex}\ldots,\hspace{-0.2ex}c_{i_1\hspace{-0.2ex}-\hspace{-0.2ex}1},c_{i_1},\hspace{-0.2ex}\dots,\hspace{-0.2ex}c_{i_1+t_1\hspace{-0.2ex}-\hspace{-0.2ex}1})
\end{align*}
are identical, so the subvector $(c_{i_1}\hspace{-0.1ex},\hspace{-0.1ex}c_{i_1\hspace{-0.1ex}+\hspace{-0.1ex}1},\hspace{-0.2ex}\dots,\hspace{-0.2ex}c_{i_1+t_1\hspace{-0.2ex}-\hspace{-0.2ex}1}\hspace{-0.1ex},\hspace{-0.1ex}c_{i_1\hspace{-0.1ex}+\hspace{-0.1ex}t_1})$ forms a run of length $t_1+1$, in contradiction to the construction of the code $ \C_3(n,\leq 2, t_1)$.
\item Case 2: If $i_2-i_1 \leq t_1$ then $i_1+t = i_1 + 2t_1 -2 \geq i_2+t_1-2$. Therefore, the first $i_2+t_1-3$ bits in the first two heads, which are subvectors
\begin{align*}
& \bfc(\delta_{ i_1},\delta_{ i_2})[1,i_2+t_1-3] \\
&=(c_1,\dots,c_{i_1-1},c_{i_1+1},\dots,c_{i_2-1},c_{i_2+1},\dots,c_{i_2+t_1-1}),\\
&\bfc(\delta_{ i_1+t},\delta_{ i_2+t})[1,i_2+t_1-3] \\
&=(c_1,\dots,c_{i_1-1},c_{i_1},\dots,c_{i_2-2},c_{i_2-1},\dots,c_{i_2+t_1-3}),
\end{align*}
are identical, which implies that $(c_{i_2-1},c_{i_2},\dots,c_{i_2+t_1-1})$ is a subvector of length $t_1+1$ with period 2, again in contradiction to the construction of the code $ \C_3(n,\leq 2, t_1)$.
\end{itemize}

Let $j_1$ be the leftmost index that $\bfc(\delta_{ i_1},\delta_{ i_2})$ and $\bfc(\delta_{ i_1+t},\delta_{ i_2+t})$ differ.
Similarly to the argument in the proof of Theorem \ref{thm1.1}, such an index exists and we can obtain the vector $\bfc(\delta_{i_2})$ by concatenating the first $j_1$ bits in $\bfc(\delta_{ i_1+t},\delta_{ i_2+t})$ and the last $n-1-j_1$ bits in $\bfc(\delta_{ i_1},\delta_{ i_2})$, i.e., $\bfc(\delta_{i_2}) = \bfc(\delta_{ i_1+t},\delta_{ i_2+t})[1,j_1] \circ \bfc(\delta_{ i_1},\delta_{ i_2})[j_1,n-2]$.
\end{IEEEproof}

The next example demonstrates the decoding procedure presented in Theorem~\ref{multideletion:thm1}.\vspace{-1.5ex}
\begin{example}
Let $n=11,t_1=3,t=4$ and the stored codeword is $\bfc=(0,0,1,1,0,1,1,0,1,1,1)\in\C_2(11,\leq 2,3).$
Assume that the outputs from the three heads are:
\begin{align*}
\textbf{Head 1: } \bfc(\delta_1,\delta_3)&=(0,1,0,1,1,0,1,1,1),\\
\textbf{Head 2: } \bfc(\delta_5,\delta_7)&=(0,0,1,1,1,0,1,1,1),\\
\textbf{Head 3: } \bfc(\delta_9,\delta_{11})&=(0,0,1,1,0,1,1,0,1).
\end{align*}
By comparing the outputs from the first two heads, we see that $j_1=2$ is the leftmost index that $\bfc(\delta_1,\delta_3)$ and $\bfc(\delta_5,\delta_7)$ differ. Hence, we can obtain the vector
\[\hspace{-0.3ex}\bfc(\delta_3)\hspace{-0.3ex}=\hspace{-0.3ex}\bfc(\delta_5,\delta_7)[1,2]\circ \bfc(\delta_1,\delta_3)[2,9]\hspace{-0.3ex}=\hspace{-0.3ex}(0,0,1,0,1,1,0,1,1,1).\]
Similarly, we find the leftmost index that $\bfc(\delta_5,\delta_7)$ and $\bfc(\delta_9,\delta_{11})$ differ which is $j_2=5$ and obtain the vector
\[\hspace{-0.3ex}\bfc(\delta_7)\hspace{-0.3ex}=\hspace{-0.5ex}\bfc(\delta_9,\delta_{11})[1,5]\circ\bfc(\delta_5,\delta_7)[5,9]\hspace{-0.3ex}=\hspace{-0.3ex}(0,0,1,1,0,1,0,1,1,1).\]
Now, we can recover the original codeword by finding $j_3=4$ as the leftmost index that $\bfc(\delta_3)$ and $\bfc(\delta_7)$ differ and recover the stored codeword $\bfc$ to be
\[\bfc=\bfc(\delta_7)[1,4] \circ \bfc(\delta_3)[4,10]=(0,0,1,1,0,1,1,0,1,1,1).\vspace{-2ex}\] 
\end{example}\vspace{-4ex}

Based on the cardinality result on the code $\C_3(n,\leq 2,t_1)$ from Section~\ref{sec:burst} we conclude with the following corollary.
\begin{corollary}
There exists a three-head double-deletion-correcting code with at most a single bit of redundancy when the distance between adjacent heads is at least $t=2(\ceil{\log(n)}+2)$.
\end{corollary}

The idea in the proof of Theorem~\ref{multideletion:thm1} was to use every two pairs of adjacent heads in order to correct the first deletion in the first head in each pair. It turns that this basic procedure is all we need in order to generalize the construction to $m$-head $d$-deletion-correcting codes. Due to the lack of space we only state here the results of these constructions.
\begin{theorem}\label{th:multiple_less_heads}
Let $\C$ be a $(d-m+1)$-deletion-correcting code, where $m\leq d$. Then, the code $\C\cap \C_3(n,\leq d, t_1)$ is an $m$-head $d$-deletion-correcting code where the distance between adjacent heads is $t \geq d t_1 - d(d+1)/2 +1$. In particular under this setup:
\begin{enumerate}
\item There exists a $(d+1)$-head $d$-deletion-correcting code with at most a single bit of redundancy
\item There exists a $d$-head $d$-deletion-correcting code with redundancy at most $\ceil{\log(n+1)}+1$.
\end{enumerate}
\end{theorem}

Lastly, we report on our results for the other cases solved in this work, which we could not include their details.\vspace{-1ex}
\begin{theorem}
\begin{enumerate}
\item The code $\C_1(n,1, t)$ can correct $d$ bursts of sticky insertions each of length at most $t-1$ using $d+1$ heads while the distance between adjacent heads is at least $t$. Specifically, for $t=\ceil{\log(n)}+1$, the redundancy of the code is approximately $0.36$ bits.
\item The code $\C_1(n,1,t)$ is a two-head single-position-error-correcting code when the distance between two heads is at least $t$.
\item The code $\C_3(n,\leq 2, t_1)$ is a three-head two-position-error-correcting code when the distance between adjacent heads is at least  $t=3t_1-2.$
\end{enumerate}
\end{theorem}
\vspace{-1ex}

\end{document}